\documentclass{article}
\usepackage[tbtags]{amsmath}
\usepackage{amsmath}
\usepackage[portrait, margin=1in]{geometry}
\usepackage{amssymb}
\usepackage{booktabs}
\usepackage{tikz}
\usetikzlibrary{positioning}
\usepackage{bbm}
\usepackage{authblk}
\usepackage{float}
\usepackage[toc,page]{appendix}
\usepackage{bm}
\usepackage[bottom]{footmisc}
\usepackage{tikz}
\usetikzlibrary{bayesnet}
\usetikzlibrary{arrows}
\usepackage{color}
\usepackage{graphicx}
\usepackage{caption}
\usepackage{subcaption}
\usetikzlibrary{backgrounds}
\usepackage{upgreek}
\usepackage{graphicx}
\usepackage{mathtools}
\usepackage[T1]{fontenc}
\usepackage{lipsum}   % for filler text
\usepackage{setspace} % for \onehalfspacing and \singlespacing macros
\usepackage{algorithm}
\usepackage{algpseudocode}
\usetikzlibrary{positioning}
\usepackage[utf8]{inputenc}
\usepackage{algorithm,algcompatible,amssymb,amsmath}
\usepackage{algorithm}
\usepackage{algpseudocode}
\renewcommand{\COMMENT}[2][.5\linewidth]{%
  \leavevmode\hfill\makebox[#1][l]{//~#2}}
\algnewcommand\algorithmicto{\textbf{to}}
\algnewcommand\RETURN{\State \textbf{return} }

\title{Multi-Industry Simplex :\\ A Probabilistic Extension of GICS}
\author{Maksim Papenkov$^{1,2,}$\footnote{Corresponding Author : mp3827@columbia.edu}\;, Chris Meredith$^{1,3}$, Claire Noel$^1$, \\ Jai Padalkar$^1$, Temple Hendrickson$^1$, Daniel Nitiutomo$^1$, $\&$ Thomas Farrell$^1$}

\date{%
    $^1$O'Shaughnessy Asset Management\\%
    $^2$Columbia University, Department of Computer Science\\%
    $^3$Cornell University, Johnson College of Business\\[5ex]%
    \today
}

\begin{document}

\maketitle
\vspace{-0.6cm}
\begin{abstract}  
    Accurate industry classification is a critical tool for many asset management applications. While the current industry gold-standard GICS (Global Industry Classification Standard) has proven to be reliable and robust in many settings, it has limitations that cannot be ignored. Fundamentally, GICS is a single-industry model, in which every firm is assigned to exactly one group - regardless of how diversified that firm may be. This approach breaks down for large conglomerates like Amazon, which have risk exposure spread out across multiple sectors. We attempt to overcome these limitations by developing MIS (Multi-Industry Simplex), a probabilistic model that can flexibly assign a firm to as many industries as can be supported by the data. In particular, we utilize topic modeling, an natural language processing approach that utilizes business descriptions to extract and identify corresponding industries. Each identified industry comes with a relevance probability, allowing for high interpretability and easy auditing, circumventing the black-box nature of alternative machine learning approaches. We describe this model in detail and provide two use-cases that are relevant to asset management - thematic portfolios and nearest neighbor identification. While our approach has limitations of its own, we demonstrate the viability of probabilistic industry classification and hope to inspire future research in this field. 
\end{abstract}

\vspace{0.2cm}

\newpage
\section{Research Motivation}

\subsection{Problem}

Is Amazon a bookstore? Perhaps in 1998 it was, but even then - it wasn't just a bookstore, rather it was a \emph{digital} bookstore (making it \emph{both} a retailer, and a fledgling internet company). But what about now? Amazon is still technically a bookstore, but it's also a cloud computing platform, a film studio, a grocer, and so much more. Amazon is gradually evolving into a conglomerate of conglomerates, leaving investors with a predicament - if they are holding a stock with such heterogeneous risk exposure, how can they manage it within a complex portfolio? 

\subsection{Existing Solution (and its Limitations)}

For such a task, most practitioners turn to \textbf{GICS (Global Industry Classification Standard)}, which assigns each firm to exactly \emph{one} industry - generally determined by the sub-business of a firm that generates the highest revenue \cite{GICS}). The more diversified a conglomerate becomes however, the less its total risk can be accurately attributed to just a single sub-business. When asset managers operate off of such incomplete information, they expose their portfolio to \textbf{misrepresentation risk}. 

If we label Amazon as a ``retail'' stock - we introduce \emph{misrepresentation risk}. If we then instead label it as a ``tech'' stock, we then just introduce alternative \emph{misrepresentation risk} - there's no way around it. As large-cap firms continue to diversify, misrepresentation risk becomes increasingly more difficult for investors to ignore - and a critical demand for a multi-industry classification tool reveals itself. While GICS has proven to be a robust and reliable system for many decades, the ubiquity of multi-sector diversification for large-cap firms requires an honest consideration of GICS' key limitations :  
\vspace{-0.3cm}
\begin{itemize}
    \item \textbf{One-Dimensional : } A firm can only belong to \emph{one} industry. 
    \item \textbf{Statically-Defined :} Industry definitions are rarely updated and may become \emph{outdated} over time.  
    \item \textbf{Opaque : } Firms are \emph{manually} classified (by committee) after an IPO, with rare label re-assignments. 
\end{itemize}
\vspace{-0.3cm}
Asset managers need a better tool ; we provide one by leveraging recent innovations in artificial intelligence. 

\subsection{New Solution (and its Value Proposition)}

We introduce the \textbf{Multi-Industry Simplex (MIS)} as a probabilistic extension of GICS. Rather than assign each firm to exactly \emph{one} industry, we now assign it to \emph{multiple} industries - in proportion to their relevance (estimated by a model). MIS provides a rigorously-constructed, data-oriented solution that provides asset managers with a better tool for representing risk, with the following specific benefits : 
\vspace{-0.3cm}
\begin{itemize}
    \item \textbf{Multi-Dimensional :} A firm can belong to \emph{multiple} industries. 
    \item \textbf{Dynamically-Defined :} Industry definitions are flexible and can easily \emph{evolve} with new data. 
    \item \textbf{Internally-Consistent :} Firms are classified \emph{simultaneously}, guaranteeing model homogeneity.
\end{itemize}
\vspace{-0.3cm}
MIS is a text-based classification tool that leverages \emph{natural language processing} to identify the key industries that a firm is involved in. We discuss the input data in section 2, and describe the model architecture in section 3. While our method has limitations of its own, we argue that a probabilistic framework is the natural evolution for industry classification systems - which can provide significant positive impact to various asset management applications.

\newpage
\section{Principals for Robust Probabilistic Industry Classification}
\vspace{-0.15cm}
Before we dive into the methodology for MIS, we first establish heuristics for robust probabilistic industry classification such that our model  best satisfies investor needs by identifying a critical requirement :
\vspace{-0.3cm}
\begin{itemize}
    \item \textbf{Explainable Fail States :} data is noisy, markets are wild, and with full certainty - at some point every single model will invariably fail. Investors generally understand this, and reluctantly accept it as a necessary price for participation. What investors generally do \emph{not} tolerate however, is an \emph{unexplainable} fail state in which the asset manager is not able to attribute model failure to any particular model defect that can potentially be circumvented with further research and model iteration. 
\end{itemize}
\vspace{-0.3cm}
We believe that this is the single most significant barrier to the utilization of advanced artificial intelligence techniques in asset management. Though a model may be correct $99\%$ of the time, if an asset manager is unable to explain the $1\%$ of events in which they lose a significant amount of money for a client, they are at a high risk of losing investor confidence, which ultimately hurts their business. 

\subsection{Model Risk Management}
\vspace{-0.15cm}
MIS, along with other probabilistic industry classification methods, are not simply models for the sake of prediction - rather they are proper \emph{risk management} tools, and must be held to the highest standard of quality, as model failure can lead to the very rapid erosion of an investor's wealth (e.g. Long Term Capital Management). To manage model risk, we require three specific criteria : 
\vspace{-0.3cm}
\begin{itemize}
    \item \textbf{Numerical Stability :} small perturbations in input data must yield negligible prediction variation. 
    \item \textbf{Replicability :} model predictions should be approximately invariant to initialization. 
    \item \textbf{Interpretability :} each model parameter must be plainly relatable to a real-world attribute (later on in the paper we'll see that every single parameter of MIS can be plainly interpreted as a conditional probability of some object belong to a category). 
\end{itemize}
\vspace{-0.3cm}
Many modern innovations in regularization reliably ensure that most models satisfy the first two criteria, but the third is far more difficult to obtain, particularly for \emph{black-box} models such as neural networks, which have been utilized for probabilistic industry classification in innovative and exciting papers such as \cite{blackbox1, blackbox2, blackbox3, blackbox4}. 

\vspace{-0.15cm}
\subsection{Practical Risk of Black-Box Models}
\vspace{-0.15cm}
While neural networks (such as large language models like GPT \cite{gpt4} and LLaMa \cite{llama}) have revolutionized the ways in which we utilize natural language data for real-world applications, they often present a significant risk : \emph{unexplainable fail states}. It is widely understood that this is simply the cost of high-precision modeling, but it is important to recognize that this risk is much easier to ignore in applications where the implication of model failure is much less severe - such as text summarization, chatbots, and machine translation. We argue that many investors are simply not willing to tolerate such risks when the cost of failure is a significant loss in their wealth. 

\subsection{Relative Benefit of Clear-Box Models}
\vspace{-0.15cm}
Though in many regards the architecture of MIS may be considered more \emph{naive} than an LLM (as we'll demonstrate in later sections), we argue that the mechanical transparency of the underlying probability model provides an asset manager with a clear framework to understand and explain why a model fails - which would then allow for quick iteration to address any issues that are discovered. In the context of risk management, model simplicity is often an asset, not a liability. 

\newpage
\section{Data}

At a high-level, MIS extracts high-importance keywords from a large collection of text to identify the most relevant industries corresponding to a firm. As with any data-oriented endeavor, the reliability of this industry classification method is highly dependent on the quality of data sourcing and the robustness of the text pre-processing methodology. \emph{Better curated input data invariably leads to a more useful model.} 

\vspace{-0.3cm}
\subsection{Sourcing}

To improve the probability of correctly associating keywords with relevant industries, we require comprehensive documents with high word-counts to provide sufficient evidence for classification. In practice, it is recommended to pool together text data from multiple sources to ensure coverage of all relevant industries that a firm is involved in. Examples of appropriate texts include : 
\vspace{-0.3cm}
\begin{itemize}
    \item \textbf{10-K Business Descriptions :} self-disclosed descriptions of high-relevance business ventures. 
    \item \textbf{Analyst-Written Business Descriptions :} third-party descriptions of business ventures. 
    \item \textbf{Earnings Call Transcripts :} discussions on existing and emerging ventures of a firm. 
\end{itemize}
\vspace{-0.3cm}
Beyond this, we can also supplement our dataset with news headlines and social media feeds - though such data sources provide less structure and require additional considerations for stable integration. In practice, we generally concatenate multiple separate documents into a single object. As we'll demonstrate later, the relative weighting of a document in the model is proportionate to the number of words : a twenty-paragraph business description will have much more influence on classification than a ten-word news headline. 

\textbf{Fundamental Data Assumptions :} all documents are \emph{comprehensive} (they cover \emph{all} non-trivial business ventures of a firm) and are \emph{proportionately-representative} (business-ventures are mentioned in proportion to their relevance to a firm; if something is discussed in greater detail - it must have higher relevance). We can reasonably assume that documents such as 10-K business descriptions satisfy these assumptions by default, but for exotic text datasets like social media feeds this must be verified by the practitioner. 

\subsection{Bag-of-Words Representation (BOW)}

In principal, MIS identifies industries by counting keywords - which thus requires keywords to be easily extractable from our text. For this, we require the \textbf{bag-of-words (BOW)} representation, which is a permutation-invariant multiset of words. Simply, we care about the frequency with which a word  appears in a document, but not the particular ordering. Consider the example below : 
\begin{equation*}
\text{``Amazon sells many books.''} \Rightarrow 
    \begin{cases}
        \{\text{\;``Amazon''\;,\; ``sells''\;,\; ``many''\;,\; ``books.''\;}\} 
        \\[5pt]
        \{\text{\;``books.''\;,\;``many''\;,\;``sells''\;,\; ``Amazon''\;}\}
        \\[5pt]
        \{\text{\;``many''\;,\;``Amazon''\;,\;``books.''\;,\; ``sells''\;}\}
    \end{cases}
\end{equation*}
The primary limitation of the BOW representation is that it is unable to utilize within-document context to extract meaning from text, unlike more sophisticated embedding schemes utilized by modern neural-net based NLP models like GPT \cite{gpt4}. 

\textbf{Fundamental BOW Limitation :} \emph{induced semantic ambiguity}, which is the inability to distinguish a particular usage of a word in the absence of context. To make this less abstract, consider for example the word ``tissue'' without additional descriptive information. On its own - it is unclear whether this word corresponds to a materials industry (as in ``paper tissue'') or to a healthcare industry (as in ``human tissue''). Thus, we are required to apply text pre-processing measures to perform semantic disambiguation. 

\newpage
\subsection{Text Pre-Processing}\label{text_pre_processing}

To mitigate the influence of noise and ambiguity on industry classification, we must apply various \emph{text pre-processing} techniques to construct a \textbf{semantically-unambiguous BOW}, which is simply a collection of words such that each word is logically associated with only \emph{one} industry. 

The goal of text pre-processing is to homogenize our input data, which reduces the complexity of our vocabulary and thus reduces the dimensionality of our model. This includes the following set of tools : 
\vspace{-0.3cm}
\begin{itemize}
    \item \textbf{Normalization} : remove punctuation and convert all letters to lower-case. 
    \item \textbf{Stemming} : standardize a word by reducing it to its root (\;``movies'' $\Rightarrow$ ``movie''\;).
    \item \textbf{N-Gram Construction} : form compound words (\;``machine'' + ``learning'' $\Rightarrow$ ``machine-learning''\;).
    \item \textbf{Stopword Removal} : eliminate non-descriptive words (\;``and''\;).
\end{itemize}
\vspace{-0.3cm}
Consider the following example which fully pre-processes an entire sentence : 
\begin{equation*}   \text{``Amazon provides machine learning tools and makes movies."} \Rightarrow \big\{\text{\;``machine-learning''\;,\;``movie''\;}\big\}
\end{equation*}
Further, we can also apply \textbf{lemmatization} which maps a set of words to a single keyword that represents aggregate meaning. These can either be synonyms or sub-cases of a general term. For example : 
\begin{equation*}
    \begin{matrix}
        \text{``clean-energy''}
        \\[2.5pt]
        \text{``green-energy''}
        \\[2.5pt]
        \text{``green-power''}
    \end{matrix}\;\;
    \Bigg\}
    \Rightarrow
    \text{``renewable-energy''}
\end{equation*}
When building a text pre-processor for a large collection of documents, it is helpful to encode these various steps in a \textbf{semantic tree} which summarizes how all of the various tools operate together to homogenize our vocabulary. Consider the following illustrative example, which constructs the keyword ``movie'' : 
\begin{figure}[h!]
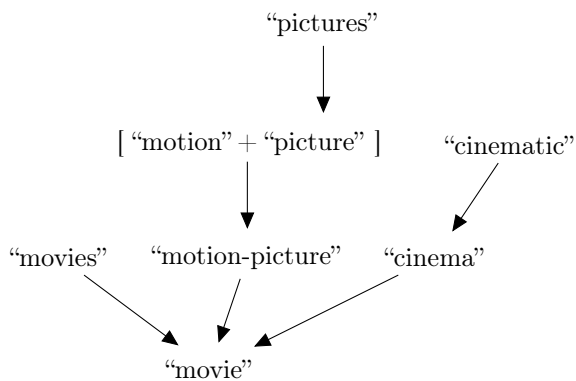

\centering
  \tikz{

  \node (movie) {``movie''};
  \node[above=of movie, xshift=-2.0cm] (movies) {``movies''};
  \node[above=of movie, xshift=+0.5cm, yshift=-0.05cm] (motion_picture) {``motion-picture''};
  \node[above=of movie, xshift=+3.0cm] (cinema) {``cinema''};
  \node[above=of cinema, xshift=+1.0cm] (cinematic) {``cinematic''};
  \node[above=of motion_picture] (plus) {+};
  \node[left=of plus, xshift=+1.2cm] (motion) {[ ``motion''};
  \node[right=of plus, xshift=-1.15cm] (picture) {``picture'' ]};
  \node[above=of picture] (pictures) {``pictures''};

  \draw[->, shorten >=3pt] (movies)  -- (movie);
  \draw[->, shorten >=3pt] (motion_picture)   -- (movie);
  \draw[->, shorten >=3pt] (cinema)   -- (movie);
  \draw[->, shorten >=3pt] (plus) -- (motion_picture);
  \draw[->, shorten >=3pt] (pictures) -- (picture);
  \draw[->, shorten >=3pt] (cinematic) -- (cinema);

 }
 \caption{Partial Semantic-Tree for Keyword = ``movie''}
\end{figure}\label{semantic_tree}

\vspace{-0.5cm}
For this example we construct a bigram (``motion'' + ``picture'' $\Rightarrow$ ``motion-picture''), we apply stemming (``cinematic'' $\Rightarrow$ ``cinema'') and we apply lemmatization (``cinema'' $\Rightarrow$ ``movie''). In practice, a full semantic tree may be vast with hundreds of nodes that account for subtle nuances in language, though regardless of scale - the logic doesn't need to be more complicated that what we demonstrate above. 

We recommend that such semantic trees are built by hand, to ensure validity and thoroughness. We emphasize that \emph{a domain expert's ability to manipulate the input text governs the nature of the industries that emerge from the probabilistic model}. Language is subjective, and a practitioner's biases are reflected in the model - though we argue that this isn't a bug, It's a feature. The goal of MIS is not to blindly adhere to a black-box modelling approach, but rather to leverage human expertise to guide the model towards a solution that best meets an asset manager's specific needs. 

\newpage
\section{Model}

\vspace{-0.2cm}
The modeling framework for MIS utilizes \textbf{Bayes Theorem}, which incorporates information contained in dataset $\mathbf X$ and prior parameter $\boldsymbol\theta$ to estimate a \emph{posterior} distribution as : 
\begin{align}
    \underbrace{\mathbb P\big(\boldsymbol\theta\mid \mathbf X \big)}_\text{posterior} \propto \underbrace{\mathbb P\big(\mathbf X \mid\boldsymbol\theta\big)}_\text{likelihood} \;\; \underbrace{\mathbb P\big(\boldsymbol\theta\big)}_\text{prior}
\end{align}
For the remainder of this section, we assume that the reader is familiar with basic Bayesian learning theory. 

\subsection{Mixture Model}

\vspace{-0.2cm}
We motivate the architecture of \emph{mixed-membership} models by first describing \emph{mixture} models, which represent an observation as a distribution across several groups\footnote{Mixture models can be thought of as probabilistic generalizations of the classical machine learning tool \emph{K-Means Clustering}.}. The core machinery of a mixture-model is the \textbf{categorical distribution}, which represents random variable $\mathbf z$ as of one of $K$-many possible values based on probability distribution $\boldsymbol\theta \in \triangle^K$ (which we call a ``simplex'') : 
\begin{equation}
\mathbf z \sim \text{Categorical}_K\big(\mathbf z \mid \boldsymbol\theta\big) : 
    \begin{cases}
      \mathbb P\big(z=1\mid\boldsymbol\theta\big) = \theta_1
      \\[5pt]
      \dots
      \\[5pt]
      \mathbb P\big(z=K\mid\boldsymbol\theta\big) = \theta_K
    \end{cases}
\end{equation}
As an illustrative example, with $K=2$ we can represent the fairness of a coin, and with $K=6$ we can represent the fairness of dice. For our purposes, we'll use this framework to model the probability that a firm belongs to one of $K$-many possible industries. 

To make this a proper probabilistic model, we parameterize the randomness of parameter $\boldsymbol\theta$ with a \textbf{Dirichlet distribution}, which captures the \emph{uncertainty of our uncertainty} as : 
\begin{equation}
    \boldsymbol\theta \sim \text{Dirichlet}_K\big(\boldsymbol\theta \mid \boldsymbol\beta\big) = \frac{\Gamma\big(\sum_{i=1}^K\beta_i\big)}{\prod_{i=1}^K \Gamma \big(\beta_i\big)}\cdot \prod_{i=1}^K \theta_i ^{\beta_i - 1}
\end{equation}
We combine these components to construct a \textbf{mixture model} defined by the following generative process, where $N$ is the total number of firms (each denoted as $\mathbf y_n$ to avoid confusion with $\mathbf x_n$ in the next section). 
\begin{align}
    \mathbf y_n &\sim \mathbb P\big(\mathbf y_n \mid \boldsymbol\mu_{ z_n}\big)
    \\
    \mathbf z_n &\sim \text{Categorical}_K\big(\mathbf z_n \mid \boldsymbol\theta\big)
    \\
    \boldsymbol\theta &\sim \text{Dirichlet}_K\big(\boldsymbol\theta \mid \boldsymbol\beta\big)
\end{align}
To be clear, $\mathbf z_n \in \{1, 2, \dots, K\}$ is a group index, so observation $\mathbf y_n$ simply belongs to the group-$\mathbf z_n$. Here $\boldsymbol\mu_{z_n}$ is a parameter that describes the nature of that particular group, though we leave it abstract for generality. We summarize the full architecture with the following graphical model :  
\begin{figure}[h!]
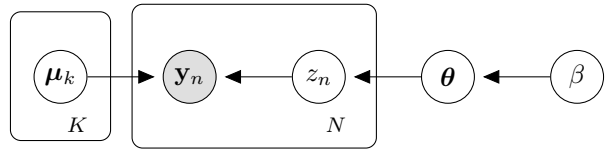

\centering
  \tikz{

 \node[obs] (y) {$\mathbf y_n$};

 \node[latent,right=of y] (z)      {$ z_n$}; 
 \node[latent,left=of y] (mu)      {$\boldsymbol\mu_k$};
 \node[latent,right=of z] (theta) {$\boldsymbol\theta$};
 \node[latent, right=of theta] (beta) {$\beta$};
 
 \plate [inner sep=.4cm,yshift=.2cm] {plate1} {(y)(z)} {$N$}; 
 \plate [inner sep=.3cm,yshift=.2cm] {plate1} {(mu)} {$K$}; 

 \draw[->, shorten >=3pt] (z) -- (y);
 \draw[->, shorten >=3pt] (mu) -- (y);
 \draw[->, shorten >=3pt] (theta) -- (z);
 \draw[->, shorten >=3pt] (beta) -- (theta);

 }
 \caption{Mixture Model}
\end{figure}
\vspace{-0.5cm}

Once this model is fit, we can then estimate the probability of group membership for a particular observation $\mathbb P(\mathbf z_n \mid \mathbf x_n)$ using Bayes Theorem, which may yield something like the following (assuming $K=2$) : 
\begin{align*}
    \mathbb P\big(\mathbf z_n \mid \mathbf y_n\big) = \big[0.95, 0.05\big] \iff \text{95\% probability $\mathbf y_n$ belongs to group-1, 5\% otherwise}
\end{align*}
We are now equipped to generalize to a more sophisticated model architecture. 

\newpage
\subsection{Mixed-Membership Model (Latent Dirichlet Allocation)}

MIS utilizes two-levels of mixture-models by positing the following architecture : 
\vspace{-0.3cm}
\begin{itemize}
    \item Each firm is an \emph{industry-mixture}. 
    \item Each industry is a \emph{word-mixture}. 
\end{itemize}
\vspace{-0.3cm}
A \emph{mixed-membership model} is used for data in which each observation is a collection of elements (in this case a BOW) such that each element belongs to a group. Mixed-membership models have prolific applications to various fields including document classification \cite{blei2003}, social network analysis \cite{cha2012}, and genetics \cite{pritchard2000}. When mixed-membership models are applied to text, they are commonly referred to as ``topic models'', the simplest of which is \emph{Latent Dirichlet Allocation (LDA)}. Some previous work has been done in the field of applying LDA to business descriptions \cite{fang2013}, but it is still a nascent area of study. 

Before diving into the math, we provide a high-level overview of the MIS flow of information as :
\begin{figure}[h!]
\centering
\begin{tikzpicture}
 
    \node[rectangle,draw, align=left] (business_description) at (0,0) {``Amazon is a conglomerate\\ firm with business ventures in\\ online retail, cloud computing,\\ digital media, as well as... ''};
    
    \node[rectangle,draw, right=of business_description, align=left] (freq_words) {``retail'' x 20\\``cloud'' x 15\\ ``movie'' x 10\\
    ...};

    \node[rectangle,draw, right=of freq_words, align=left] (topics) {\emph{E-Commerce} ($40\%$)\\\emph{Cloud Computing} ($30\%$)\\ \emph{Movies} ($20\%$)\\
    ...};

        \node[above=of business_description, yshift=-0.9cm] {Business Description};
        \node[above=of freq_words, yshift=-0.9cm] {(Clean) Bag-of-Words};
        \node[above=of topics, yshift=-0.9cm] {Industry-Mixture};

    \draw[->, shorten >=3pt] (business_description) -- (freq_words);
    \draw[->, shorten >=3pt] (freq_words) -- (topics);
 
\end{tikzpicture}
 \caption{Industry-Mixture Identification Process}
\end{figure}
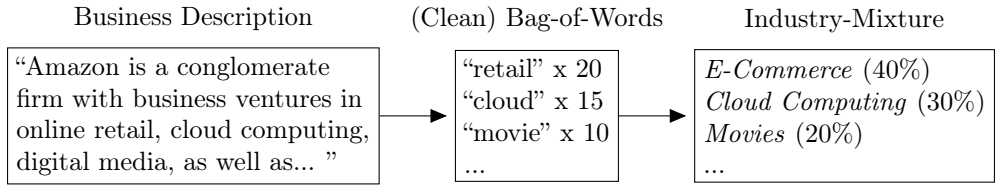
\vspace{-0.5cm}

First we begin with a raw business description, then we convert it to a clean BOW via text pre-processing, and finally the model estimates an \emph{industry-mixture}. Each industry is a \emph{word-mixture} of the following form : 
\begin{align}
\emph{E-Commerce} = \big[\;\text{``retail''} (30\%)\;,\; \text{``online''} (28\%)\;,\; \text{``supply-chain''} (25\%)\;,\; ...\big]
\end{align}
To formally define the model architecture, we require three dimensionalities : 
\vspace{-0.3cm}
\begin{itemize}
    \item $M$ : total number of firms; each business description has $N_m$-many words. 
    \item $K$ : total number of industries that are modeled (specified by the user).
    \item $V$ : total vocabulary size in pre-processed text. 
\end{itemize}
\vspace{-0.3cm}
With these, we define the following random variables : 
\vspace{-0.3cm}
\begin{itemize}
    \item $\mathbf x_m$ : the $m$-th firm's business description represented as a BOW (e.g. $\mathbf x_1 = \{\text{``bank'', ``finance'', ...}\}$).
    \item $\mathbf x_{m,n}$ : the $n$-th word in the $m$-th firm's BOW business description. 
    \item $\mathbf z_{m,n}$ : the industry-index for word $\mathbf x_{m,n}$ (such that $\mathbf z_{m,n} \in \{1, 2, \dots, K\})$. 
    \item $\boldsymbol\theta_m \in \triangle^K$ : the \textbf{industry-mixture} of the $m$-th firm.
    \item $\boldsymbol\phi_k \in \triangle^V$ : the \textbf{word-mixture} of the $k$-th industry. 
\end{itemize}
\vspace{-0.3cm}
We thus now define MIS using the \textbf{Latent Dirichlet Allocation} generative process : 
\begin{align}
    \mathbf x_{m, n} &\sim \text{Categorical}_V\big(\mathbf x_{m,n} \mid \boldsymbol\phi_{z_{m,m}}\big)
    \\
    \mathbf z_{m,n} &\sim \text{Categorical}_K\big(\mathbf z_{m,n} \mid \boldsymbol\theta_m\big)
    \\
    \boldsymbol\theta_m &\sim \mathbb P\big(\text{Industry}_k \mid \text{Firm}_m\big) = \text{Dirichlet}_K\big(\boldsymbol\theta_m\mid \boldsymbol\beta\big)
    \\
    \boldsymbol\phi_{k} &\sim \mathbb P\big(\text{Word}_k\mid \text{Industry}_m\big)=\text{Dirichlet}_V\big(\boldsymbol\phi_{k}\mid \boldsymbol\alpha\big)
\end{align}
The appeal of this architecture is that it fits both the industry-mixtures and the word-mixtures simultaneously, yielding an internally-consistent representation of a market. As each parameter in this model is simply a probability of something belonging to a particularly group, this model is fully interpretable. We discuss the hyperparameters $\boldsymbol\alpha$ and $\boldsymbol\beta$ in section 3.5.

\newpage
We summarize the full-architecture with the following graphical model : 
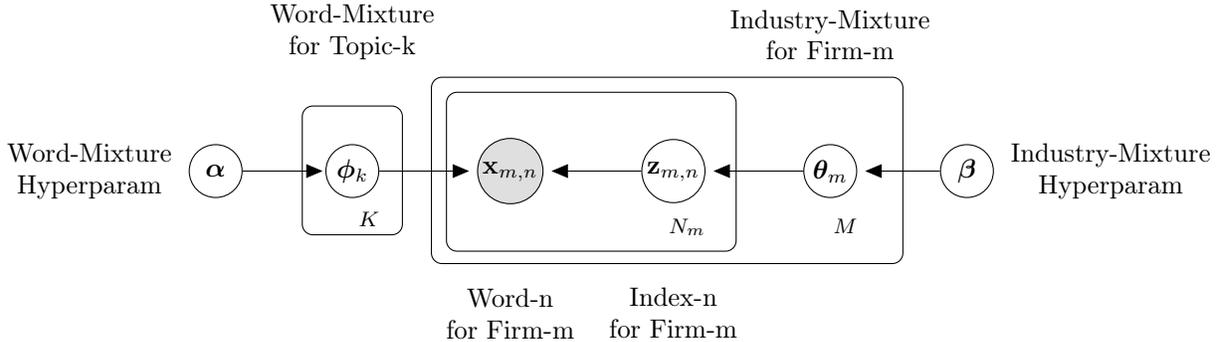
\begin{figure}[h!]
\centering
  \begin{tikzpicture}
 \node[obs                                  ] (x)     {$\mathbf x_{m,n}$};
 \node[latent, right=of x    , xshift=+0.3cm] (z)     {$\mathbf z_{m,n}$}; 
 \node[latent, left=of x     , xshift=-0.3cm] (phi)   {$\boldsymbol\phi_k$};
 \node[latent, left=of phi   , xshift=-0.1cm] (alpha) {$\boldsymbol\alpha$};
 \node[latent, right=of z    , xshift=+0.3cm] (theta) {$\boldsymbol\theta_m$};
 \node[latent, right=of theta, xshift=+0.1cm] (beta)  {$\boldsymbol\beta$};

 \node[above=of phi  , align=center] (phi_label)   {Word-Mixture\\for Topic-k};
 \node[above=of theta, align=center] (theta_label) {Industry-Mixture\\for Firm-m};
 \node[below=of x    , align=center] (x_label)     {Word-n\\ for Firm-m};
 \node[below=of z    , align=center] (z_label)     {Index-n\\ for Firm-m};
 \node[left=of alpha , align=center, xshift=0.9cm] (alpha_label) {Word-Mixture\\Hyperparam};
 \node[right=of beta , align=center, xshift=-0.9cm] (beta_label) {Industry-Mixture\\Hyperparam};
 
 \plate [inner sep=.4cm,yshift=.2cm] {plate1} {(x)(z)} {$N_m$}; 
 \plate [inner sep=.6cm,yshift=.2cm] {plate2} {(x)(z)(theta)} {$M$};
 \plate [inner sep=.3cm,yshift=.2cm] {plate3} {(phi)} {$K$};

 \draw[->, shorten >=3pt] (z) -- (x);
 \draw[->, shorten >=3pt] (phi) -- (x);
 \draw[->, shorten >=1pt] (alpha) -- (phi);
 \draw[->, shorten >=3pt] (theta) -- (z);
 \draw[->, shorten >=3pt] (beta) -- (theta);

 \end{tikzpicture}
 \caption{Mixed-Membership Model (Latent Dirichlet Allocation)}
\end{figure}
\vspace{-0.9cm}

\subsection{Model Evaluation}

A useful tool for evaluating the quality of a mixed-membership model is \emph{perplexity}, which measures goodness-of-fit. A lower perplexity indicates a better sample fit, which may provide guidance for calibrating the text pre-processor and for hyperparameter tuning. Formally, we define \textbf{perplexity} for estimated $\boldsymbol\theta$ and $\boldsymbol\phi$ as : 
\begin{equation}
    \text{Perp}\big(\boldsymbol\theta_{1:M}, \boldsymbol\phi_{1:K} \;; \mathbf x_{1:M} \big) = \exp\bigg(-\frac{\sum_{m=1}^M\log \mathbb P\big(\boldsymbol\theta_{1:M}, \boldsymbol\phi_{1:K} \;; \mathbf x_m\big)}{\sum_{m=1}^M N_m}\bigg)
\end{equation}
For this we use $\text{count}(w_{m,v})$ as the count of the $v$-th word of the vocabulary for the $m$-th firm : 
\begin{equation}
    \log \mathbb P\big(\boldsymbol\theta_{1:M}, \boldsymbol\phi_{1:K} \;; \mathbf x_m\big) = \sum_{v=1}^V \text{count}\big(w_{m,v}\big)\cdot\log\Bigg[\sum_{k=1}^K \boldsymbol\phi_{k,v} \cdot \boldsymbol\theta_{m,k}\Bigg]
\end{equation}
Perplexity is a crude measure, and at times a higher numerical perplexity may correspond to a less human-interpretable model, so care must be applied when relying on this metric. Another tool for model evaluation is \emph{coherence}\cite{simple_coherence,hinneburg_coherence}, though this is beyond the scope of this paper. We can directly compare the perplexity between two models to evaluate if one is a better fit than the other, with a lower score being more optimal. 

\subsection{Optimal Model}

To build intuitions, we present this technical aside on the nature of the joint log-likelihood for LDA : 
\begin{equation}
    \log \mathbb P\big(\boldsymbol\theta_{1:M}, \boldsymbol\phi_{1:K} \;; \mathbf x_{1:M}\big) = \sum_{k=1}^K \log\mathbb P\big(\boldsymbol\phi_k\big) +
    \sum_{m=1}^M\bigg[\log \mathbb P\big(\boldsymbol\theta_m\big) + \sum_{i=1}^{N_m}
    \big(\log\boldsymbol\theta_{m, z_{m,i}} + \log \boldsymbol\phi_{z_{m,i},w_{m,i}}\big)
    \bigg]
\end{equation}
Observe that there are precisely two mechanisms via which we can improve this function :
\vspace{-0.3cm}
\begin{itemize}
    \item For a firm, choose high-likelihood industries that maximize $\mathbb P\big(\text{industry} = z_{m,i} \mid \text{firm} = m\big)$. 
    \item For an industry, choose high-likelihood keywords that maximize $\mathbb P\big(\text{word} = w_{m,i} \mid \text{industry} = z_{m,i}\big)$.
\end{itemize}
\vspace{-0.3cm}
The key insight here is that these two objectives are \emph{mutually exclusive}, thus during inference we balance a delicate trade-off \cite{blei_mixed_membership}. By selecting fewer industries for a firm, we require them to represent a broader set of words By selecting fewer words for an industry, we require more industries to represent a single firm. 

These competing goals encourage \emph{peaky} distributions to form for both $\boldsymbol\theta$s and $\boldsymbol\phi$s, which causes both the posteriors for the industry-mixtures and the word-mixtures to appear approximately sparse. Thus most firms will be part of a small number of industries, and most industries will be defined by a small number of words. This significantly helps with interpretability in practice. 

\newpage
\subsection{Posterior Inference via Gibbs Sampling}

There exist various robust approaches for approximating the posterior of a mixed-membership model, including Gibbs-Sampling \cite{gibbs_sampler}, Expectation-Maximization \cite{blei_LDA}, and Variational Inference \cite{vi_blei}. For simplicity, here we will only focus on Gibbs Sampling for LDA, as it is mechanically transparent and provides useful intuitions for understanding how the optimal solution is obtained. 

\textbf{Gibbs Sampling} is a flavor of Monte Carlo Markov Chain (MCMC), a simulation-based technique that generates a sequence of samples that asymptotically converges to the true posterior. Here, we present the simplest form of the algorithm to develop intuitions around the underlying math \cite{lda_collapsed_gibbs, yang_stability}. 

For this algorithm, we require \emph{prior} values for $\boldsymbol\alpha^\text{prior}$ and $\boldsymbol\beta^\text{prior}$ to be specified by the user. In practice, these can either be set to constant values (producing a symmetric prior), or can be calibrated by hand in order to coerce the emergence of certain topics. Further, the user must also specify the number of industries $K$ and the total number of samples $S$. 

\begin{algorithm}
\linespread{1.4}\selectfont
  \caption{$\text{Gibbs Sampler for Latent Dirichlet Allocation }\big(\mathbf X, K, \boldsymbol\beta^\text{prior}, \boldsymbol\alpha^\text{prior}, S\big)$ :} \label{algo1}
  \begin{algorithmic}[1]
    \STATE $\{\boldsymbol \theta_{1:M}, \boldsymbol\phi_{1:K} \}^{(0)} \sim \text{Random}(\cdot\big)$ \COMMENT{Initialize word-mixtures and industry-mixtures.}
    \FOR{$s \in \{1, 2, \dots, S\}$}\COMMENT{Repeat process for $S$-many samples. }
        \FOR {$m\in\{1, 2, \dots, M\}$} \COMMENT{Per each $m$-th firm :}
            \FOR {$n \in \{1, 2, \dots, N_m\}$} \COMMENT{Per each $n$-th word for the $m$-th firm :}
            \STATE{$\mathbf z_{m,n}^{(s)} \sim \mathbb P\big(\text{Industry}_k \mid \text{Firm}_m\big)$}\COMMENT{- probabilistically assign each word to an industry.}
            \ENDFOR
        \ENDFOR
        \FOR{$k \in \{1, 2, \dots, K\}$} \COMMENT{Per each $k$-th industry: }
            \STATE {$\boldsymbol\beta_k^{(s)} \gets \text{count}\big[\mathbf z_1^{(s)}, \dots, \mathbf z_m^{(s)}\big]$} \COMMENT{- count word-$v$ assigned to industry-$k$ across all firms.}
            \STATE {$\boldsymbol\beta_k^\text{new} \gets \boldsymbol\beta_k^\text{prior} + \boldsymbol\beta_k^{(s)}$} \COMMENT{- update parameter of probability model. }
            \STATE {$\boldsymbol\phi_k^{(s)} \sim \text{Dirichlet}\big(\boldsymbol\beta_k^\text{new})$} \COMMENT{- sample a new word-mixture.}
        \ENDFOR
        \FOR {$m\in\{1,2,\dots,M\}$} \COMMENT{Per each $m$-th firm : }
            \STATE {$\boldsymbol\alpha_m^{(s)} \gets \text{count}\big[\mathbf z_m^{(s)}\big]$}\COMMENT{- count all words assigned to each $k$-th industry.}
            \STATE {$\boldsymbol\alpha_m^\text{new} \gets \boldsymbol\alpha_m^\text{prior} + \boldsymbol\alpha_m^{(s)}$} \COMMENT{- update parameter of the probability model.}
            \STATE {$\boldsymbol\theta_m^{(s)} \sim \text{Dirichlet}_K\big(\boldsymbol\alpha_m^\text{new}\big)$} \COMMENT{- sample a new industry-mixture.}
        \ENDFOR
    \ENDFOR
    \RETURN $\{\boldsymbol \theta_{1:M}, \boldsymbol\phi_{1:K} \}^{(0, \dots, S)}$\COMMENT{Return the estimated samples.}
  \end{algorithmic} 
\end{algorithm}
\vspace*{-.2cm}

In practice, we discard the first $S^*$ many of the returned samples as part of the \emph{burn-in} process, and could take the mean of the remainder as our best estimate for the optimal value. 

We highlight that fitting such a model simply amounts to \emph{counting} the number of words assigned to each industry, and then iteratively re-calibrating until we arrive at a stable solution. In general, higher values for $S$ and $S^*$ yield a better model, and in practice our only limitation to convergence is hardware. We note that there exist more elaborate extensions of this algorithm in the literature, though they are beyond the scope of this paper \cite{HDP,blei_CTM,blei_HTM,li_Pachinko}. As this architecture is clearly very simple, we emphasize that text pre-processing is the primary mechanism for influencing the optimal solution of this model.

\newpage
\section{Minimal Illustrative Example}

Consider a collection of hypothetical pre-processed BOW business descriptions for three fictional firms : 
\begin{align*}
    \color{orange}\text{Nile}  &= \big\{\text{``retail'', ``cloud'', ``movie'', ``movie'', ``retail'', ``cloud``}\big\} 
    \\[5pt]
    \color{blue}\text{Tallmart} &= \big\{\text{``retail'', ``retail'', ``store''}\big\} 
    \\[5pt]
    \color{red}\text{CloudFilms}  &= \big\{\text{``movie'', ``cinema'', ``web-app'', ``cloud''}\big\} 
\end{align*}
Suppose we fit an MIS model on this data with $K=3$ to generate the following industry representations : 
\vspace{0.2cm}
\begin{align*}
    \emph{Retail} : \phi_1 &= \mathbb P\big(\text{word}\mid \text{industry} = 1\big) = \big[\;\underset{\color{white}{\text{``cinema''}}}{0.01}, \underset{\color{white}\text{``cloud''}}{0.02}, \underset{\color{white}\text{``movie''}}{0.01}, \underset{\color{white}\text{``retail''}}{\textbf{0.61}},\underset{\color{white}\text{``store''}}{\textbf{0.33}}, \underset{\color{white}\text{``web-app''}}{0.02}\;\big]
    \\[2.5pt]
    \emph{Movie} : \phi_2 &= \mathbb P\big(\text{word}\mid \text{industry} = 2\big) =\big[\;\underset{\color{white}{\text{``cinema''}}}{\textbf {0.31}}, \underset{\color{white}\text{``cloud''}}{0.01}, \underset{\color{white}\text{``movie''}}{\textbf{0.64}}, \underset{\color{white}\text{``retail''}}{0.02},\underset{\color{white}\text{``store''}}{0.01}, \underset{\color{white}\text{``web-app''}}{0.01}\;\big]
    \\[2.5pt]
    \emph{Tech} : \phi_3 &= \mathbb P\big(\text{word} \mid \text{industry} = 3\big) = \big[\;\underset{\text{``cinema''}}{0.02}, \underset{\text{``cloud''}}{\textbf{0.45}}, \underset{\text{``movie''}}{0.02}, \underset{\text{``retail''}}{0.02},\underset{\text{``store''}}{0.01}, \underset{\text{``web-app''}}{\textbf{0.48}}\;\big] 
\end{align*}
The names of the topics are intuited from the maximal values of each distribution. \emph{Retail} is dominated by ``retail'' and ``store'', \emph{Movie} is dominated by ``movie'' and ``cinema'', and \emph{Tech} is dominated by ``cloud'' and ``web-app''. Functionally, we can ignore all of the other values, as they are arbitrarily small. 

MIS also provides us with the following \emph{industry embeddings} : 
\vspace{0.1cm}
\begin{alignat*}{2}
    \color{orange}{\theta_1} &= \mathbb P\big(\text{industry}\mid \text{firm = }\color{orange}{\text{Nile}} &&\big)=  \big[\;\underset{\color{white}{\emph{Retail}}}{\textbf{0.34}}, \underset{\color{white}{\emph{Movie}}}{\textbf{0.33}}, \underset{\color{white}{\emph{Tech}}}{\textbf{0.33}}\;\big]
    \\[2.5pt]
    \color{blue}{\theta_2} &= \mathbb P\big(\text{industry}\mid \text{firm = } \color{blue} {\text{Tallmart}}&&\big)= \big[\;\underset{\color{white}{\emph{Retail}}}{\textbf{0.98}}, \underset{\color{white}{\emph{Movie}}}{0.01}, \underset{\color{white}{\emph{Tech}}}{0.01}\;\big]
    \\[2.5pt]
    \color{red}{\theta_3} &= \mathbb P\big(\text{industry}\mid \text{firm = } \color{red}{\text{CloudFilms}}&&\big) = \big[\;\underset{\emph{Retail}}{0.01}, \underset{\emph{Movie}}{\textbf{0.66}}, \underset{\emph{Tech}}{\textbf{0.33}}\;\big]
\end{alignat*}
Within this simplified three-industry market, we see that Nile is a \emph{conglomerate} with membership dispersed across multiple industries, while Tallmart is a \emph{pure play} with membership concentrated in only one. CloudFilms is in the middle. We can visualize this data as follows : 
\begin{figure}[h]
\centering
\begin{tikzpicture}
\draw (0,0) node[anchor=north]{\begin{tabular}{c} $\phi_1$ \\ \emph{Retail}\end{tabular}}
  --  (4,0) node[anchor=north]{\begin{tabular}{c} $\phi_2$ \\ \emph{Movie} \end{tabular}}
  --  (2,3) node[anchor=south]{\begin{tabular}{c} $\phi_3$ \\ \emph{Tech} \end{tabular}}
  -- cycle;
\draw (0.4, 0.6) node[circle, color=blue  , anchor=north] {$\theta_2$};
\draw (2.0, 1.5) node[circle, color=orange, anchor=north] {$\theta_1$};
\draw (3.2, 1.2) node[circle, color=red   , anchor=north] {$\theta_3$};
\draw [dotted] (0.44, 0.35) -- (1.8, 1.11) ; 
\draw [dotted] (0.57, 0.24) -- (3.0, 0.8) ; 
\end{tikzpicture}
\caption{Minimal Example of a Multi-Industry Simplex}
\end{figure}
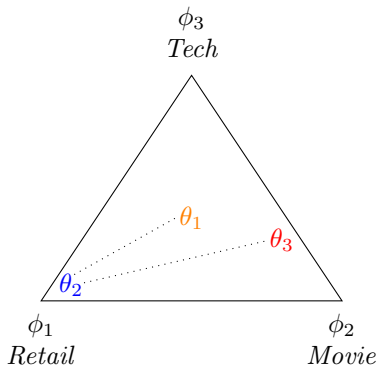

\vspace{-0.4cm}

Notice that in the above space, Tallmart ($\color{blue}{\theta_2}$) is physically closer to Nile ($\color{orange}{\theta_1}$) than it is to CloudFilms ($\color{red}{\theta_3}$). This is a mathematically rigorous way of demonstrating that Tallmart is \emph{more similar} to Nile than it is to CloudFilms, on the basis of text-similarity.

To put things into perspective, for real-world applications, we would have thousands firms in our universe, each business description would have several thousand words, and the number of industries could be over 100. At that point, visualizing the data becomes much more tricky, but all of the intuitions that we build in this low-dimensional example generalize naturally to a higher-dimensional space. 

\newpage
\section{Applications}

Utilizing a universe of over 5000 firms, we calibrate and estimate MIS with 97 industries\footnote{Due to observed instability, we elected to remove all real estate and REIT firms from our sample. This instability is primarily attributed to semantic ambiguity in the business descriptions, and exceedingly high quantities of false-positives in the model. We hope to resolve this issue in future iterations of the model.}. We emphasize that what we illustrate here is just a single implementation of what an MIS model can look like, and different choices in data sourcing and semantic tree construction will yield varying outputs. Our goal here is to motivate and justify the general methodology, rather than to fixate on a specific model instance. 

We represent the model-estimated industries with the following network, where each node corresponds to an MIS-industry. Node size is proportionate to probability of incidence  and color corresponding to the most common GICS-sector of firms within that MIS-industry. An edge between two nodes indicates that two industries have non-trivial co-occurrence, meaning that these industries are linked within the data :

\begin{figure}[h!]
    \centering
    \includegraphics[width=16.5cm]{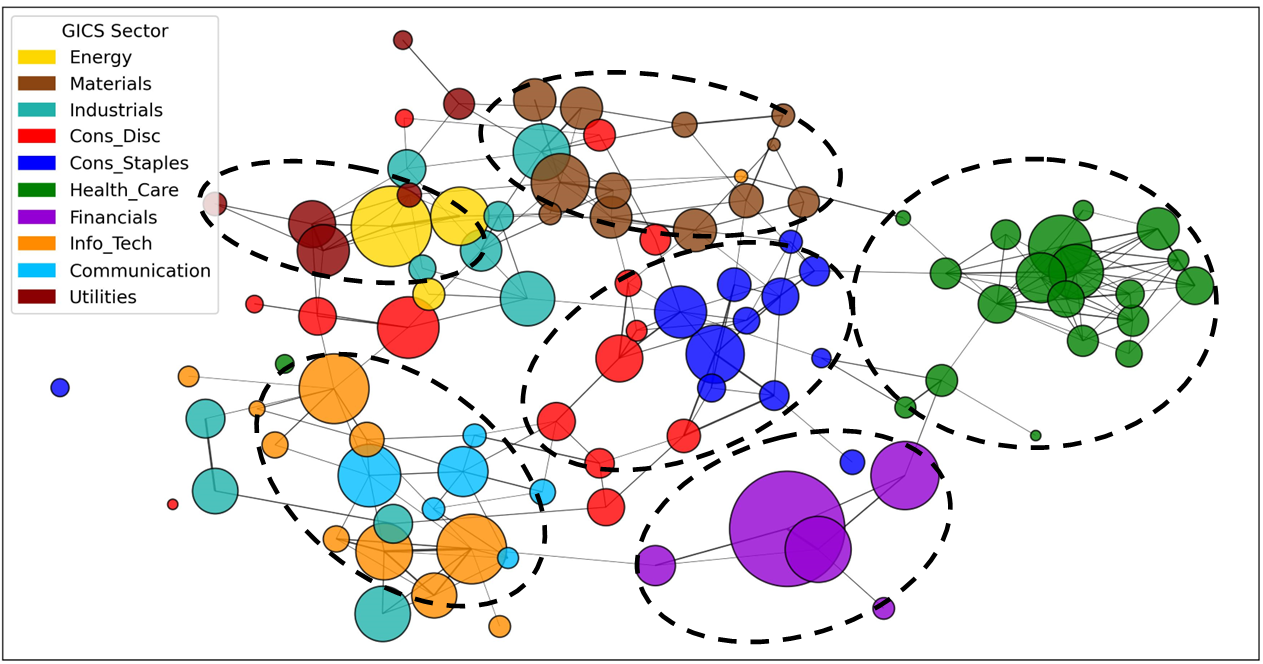}
    \caption{MIS Industry Network, Colored by Most-Associated GICS Sector}
\end{figure}

With this plot we emphasize two important points : 
\vspace{-0.3cm}
\begin{itemize}
    \item At a high-level, MIS has a lot in common with GICS, particularly in terms of \emph{global structure}. Within the plot we clearly observe a ``sector'' of health-oriented firms, a ``sector'' of finance-related firms, and so on. Here, we demonstrate that MIS doesn't attempt to reinvent the wheel. 
    \item At a more detailed-level however, MIS represents a more nuanced (and perhaps a more \emph{realistic}) representation of how industries relate to one another in a market. The abundance of edges between nodes highlights just how interconnected various industries are, which shouldn't be surprising from an economic standpoint. Thus, we argue that MIS is a \emph{better} wheel. 
\end{itemize}
\vspace{-0.3cm}

The particular MIS model instance that we show in this paper uses a large collection of private and public text datasets that are not all available before 2021, and because of this a proper historical backtest is not possible. Despite this limitation, we highlight several use-cases that are relevant to asset managers and discuss future directions of research as we strive towards improving the credibility of this methodology. 

\newpage
\subsection{Examples of MIS ``Sectors''}

To best highlight the differences between MIS and GICS, we illustrate several collections of MIS-industries that are intended to evoke the structure of GICS-sectors. 

First, consider the following commodities-related MIS industries, with the top three associated keywords : 
\begin{table}[h!]
\centering
\begin{tabular}{@{}clllll@{}}
\hline
\\
& \textbf{OIL DRILLING} & :  & ``oil-drilling'' ($82\%$) & ``gas-pipeline'' ($9\%$) & ``fracking'' ($7\%$) 
\\[5pt]
& \textbf{METAL}& :  & ``metal'' ($46\%$) & ``ferrous'' ($37\%$) & ``nonferrous'' ($14\%$)  
\\[5pt]
& \textbf{WOOD}& : & ``wood'' ($64\%$) & ``timber'' ($27\%$) & ``lumber'' ($10\%$) 
\\[5pt]
& \textbf{MEAT}& : & ``meat'' ($49\%$) & ``red-meat'' ($18\%$) & ``poultry'' ($16\%$) 
\\[5pt]
& \textbf{MARIJUANA}& : & ``marijuana'' ($54\%$) & ``cannabis'' ($44\%$) & ``t.h.c.''  ($2\%$) 
\\[5pt]
\hline
\end{tabular}
\caption{Subset of MIS Commodities Industries}
\end{table}
\vspace{-0.5cm}

The first four MIS-industries have direct analogues to GICS-industries, while \emph{marijuana} is something that is completely unrepresented in GICS. 

As a second example, consider several healthcare-oriented MIS-industries : 
\begin{table}[h!]
\centering
\begin{tabular}{@{}cllll@{}}
\hline
\\
& \textbf{ONCOLOGY :}  & ``oncology'' ($85\%$) & ``carcinoma'' ($10\%$) & ``lymphoma'' ($9\%$)  
\\[5pt]
& \textbf{MICROBIOLOGY :} & ``microbiology'' ($58\%$) & ``virology'' ($18\%$) & ``bacteriology'' ($15\%$) 
\\[5pt]
& \textbf{GENOMICS :} & ``genomics'' ($66\%$) & ``gene-editing'' ($23\%$) & ``d.n.a.'' ($10\%$) 
\\[5pt]
& \textbf{SURGERY :}  & ``surgery'' ($72\%$) & ``organ-transplant'' ($17\%$) & ``orthopedics'' ($12\%$) 
\\[5pt]
& \textbf{CARDIOLOGY :} & ``cardiology'' ($65\%$) & ``heart-attack'' ($22\%$) & ``heart''  ($11\%$) 
\\[5pt]
\hline
\end{tabular}
\caption{Subset of MIS Healthcare Industries}
\end{table}
\vspace{-0.5cm}

Unlike GICS, which simply partitions all of healthcare into broad categories like \emph{biotechnology} and \emph{pharmaceuticals}, MIS is able to leverage the specificity of medical language to identify precisely which areas of health a firm is working in, providing a much richer description of the relevant sub-industries. 

As a final example, consider some of the technology industries that MIS is able to discover : 

\begin{table}[h!]
\centering
\begin{tabular}{@{}cllll@{}}
\hline
\\
& \textbf{A.I. :} & ``artificial-intelligence'' ($60\%$) & ``computer-vision'' ($23\%$) & ``n.l.p.'' ($15\%$) 
\\[5pt]
& \textbf{STREAMING :}  & ``media-streaming'' ($49\%$) & ``television'' ($21\%$) & ``music-streaming'' ($14\%$)  
\\[5pt]
& \textbf{BLOCKCHAIN :} & ``blockchain'' ($56\%$) & ``cryptocurrency'' ($44\%$) & -
\\[5pt]
& \textbf{ROBOTICS :}  & ``robotics'' ($59\%$) & ``factory-automation'' ($31\%$) & -
\\[5pt]
& \textbf{CAD :} & ``c.a.d.'' ($52\%$) & ``3d-print'' ($48\%$) &   -
\\[5pt]
\hline
\end{tabular}
\caption{Subset of MIS Tech Industries}
\end{table}
\vspace{-0.2cm}

Here, we are again able to construct industries that are completely unrepresented in GICS (see section 5.3 for more discussion on this). As a final note here, we emphasize that the emergence of these precise industries is a function of the text sources and pre-processing choices made by the practitioner. Richer input text and more elaborate pre-processing can lead to more robust topics, and these should only be considered for illustrative purposes.

\newpage
\subsection{Nearest Neighbors}

We can utilize MIS for nearest neighbor identification, finding "most similar" firms on the basis of text. We utilize \textbf{Hellinger similarity} as a ``correlation'' measure between the descriptions of two firms : 
\begin{align}
    \text{similarity}\big(\text{firm}_i, \text{firm}_j\big) = 1 - \frac{1}{\sqrt 2} \Big\|\sqrt{\boldsymbol\theta_i} - \sqrt{\boldsymbol\theta_j}\Big\|_2
\end{align}
As an illustrative example, consider 20 Amazon neighbors with a Hellinger similarity greater than 0.25 (sorted by size). Each node represents one of these neighboring firms, where color corresponds to GICS sector and size corresponds to firm size. An edge between nodes indicates that two firms are ``neighbors''.
\vspace{-0.3cm}
\begin{figure}[h!]
    \centering
    \includegraphics[width=16cm]{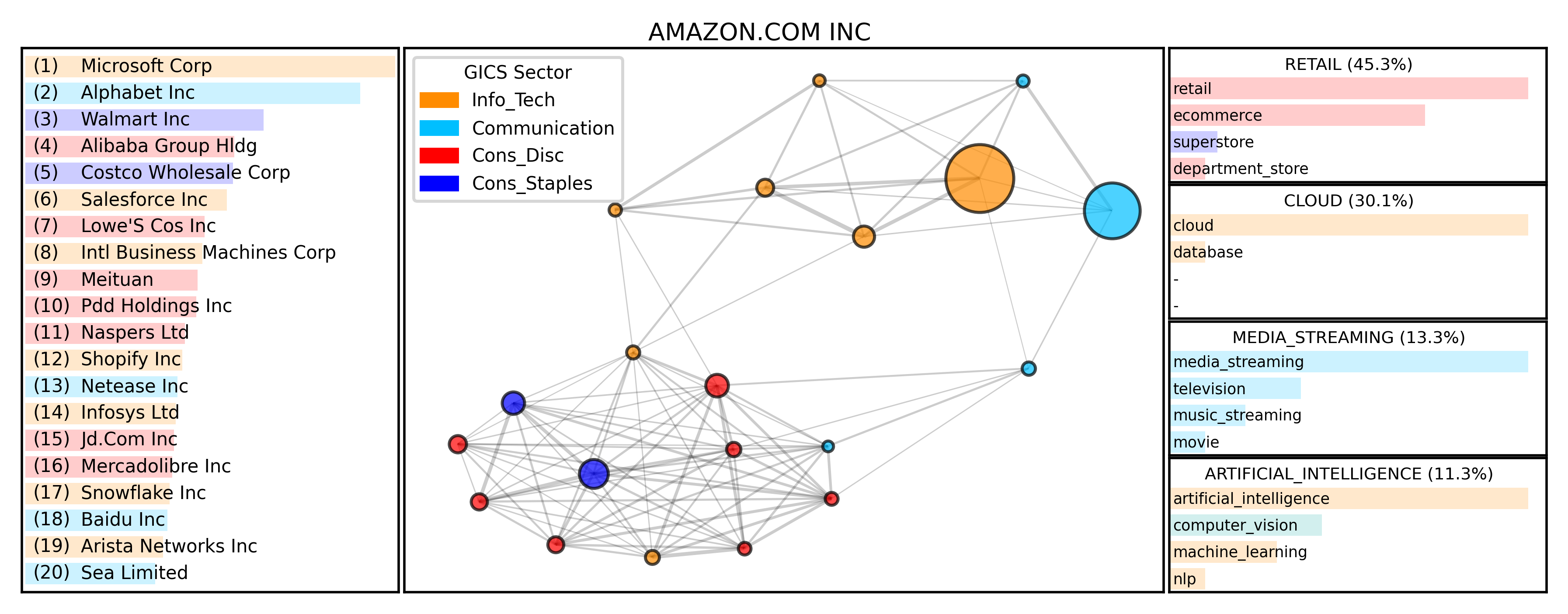}
    \caption{Nearest Neighbors for Amazon}
\end{figure}
\vspace{-0.5cm}

MIS identifies four major industries that Amazon is a part of : retail (Amazon's primary business), cloud computing (Amazon Web Services), media streaming (Amazon Prime Video), and artificial intelligence. Amazon's largest neighbors span \emph{four} GICS sectors - information technology, communications services, consumer discretionary, and consumer staples - highlighting its high level of diversification.

As a second example, consider Apple, which at the time of writing is the world's largest publicly traded firm. 
\vspace{-0.5cm}
\begin{figure}[h!]
    \centering
    \includegraphics[width=16cm]{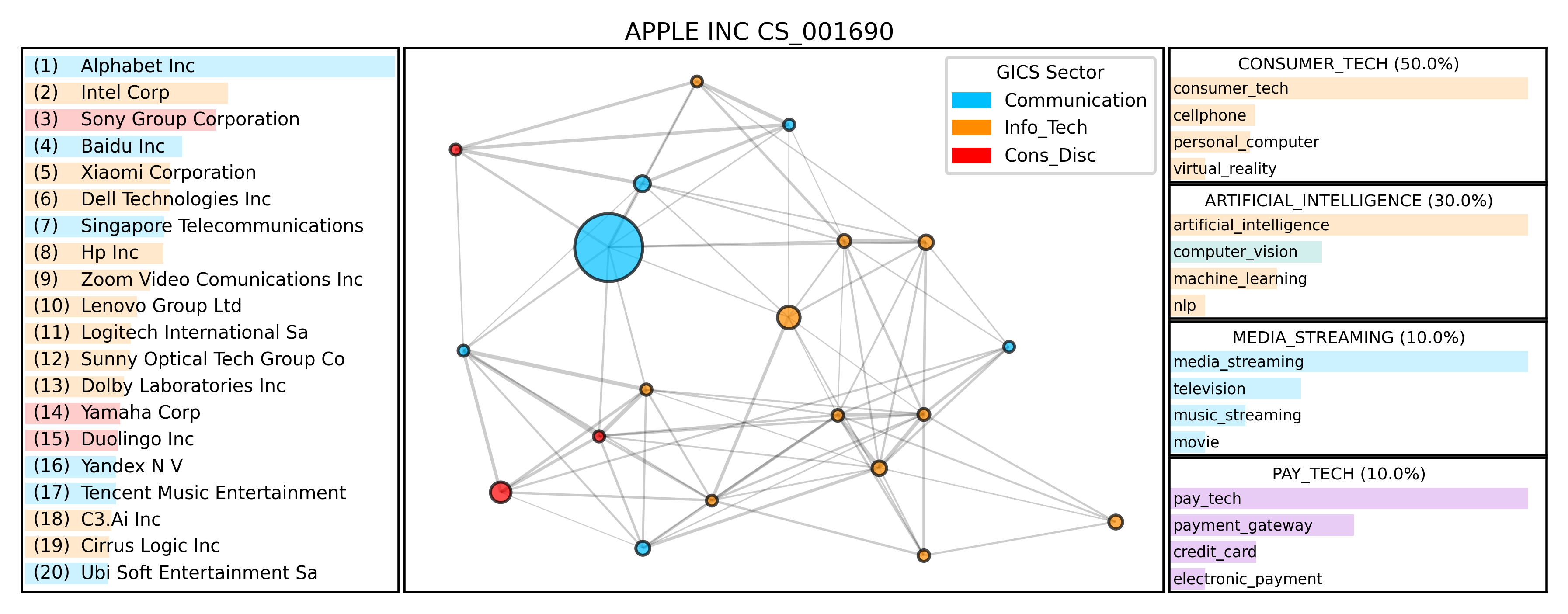}
    \caption{Nearest Neighbors for Apple}
\end{figure}
\vspace{-0.5cm}

Again, MIS is able to identify four major industries for a highly diversified firm : consumer tech (iPhone, iPad, Mac), artificial intelligence, media streaming (Apple Music, Apple TV+), and payment technology (Apple Pay). As with Amazon, the nearest neighbor list similarly represents multiple GICS sectors. 

\newpage
\subsection{Thematic Portfolios}

We can also utilize MIS for constructing thematic portfolios, which have been popular in recent years as a means of investing in innovative technologies. For illustrative example, we construct a thematic portfolio by identifying all firms with a probability of inclusion in an industry (aka ``theme'') exceeding 5$\%$. The portfolio weight is a function of firm size ($s_i$) and probability of inclusion, defined by the following formula : 
\begin{equation}
    w_i \propto \sqrt{s_i} \cdot \mathbb P\big(\text{industry}\mid\text{firm}_i\big)
\end{equation}
As an example, consider an \emph{artificial intelligence} portfolio. Again, each nodes represents a firm, where color corresponds to GICS sector and node size corresponds to firm size. The positions of the nodes are determined by tSNE \cite{tsne} such that more-similar firms are positioned more closely together.
\vspace{-0.3cm}
\begin{figure}[h!]
    \centering
    \includegraphics[width=16cm]{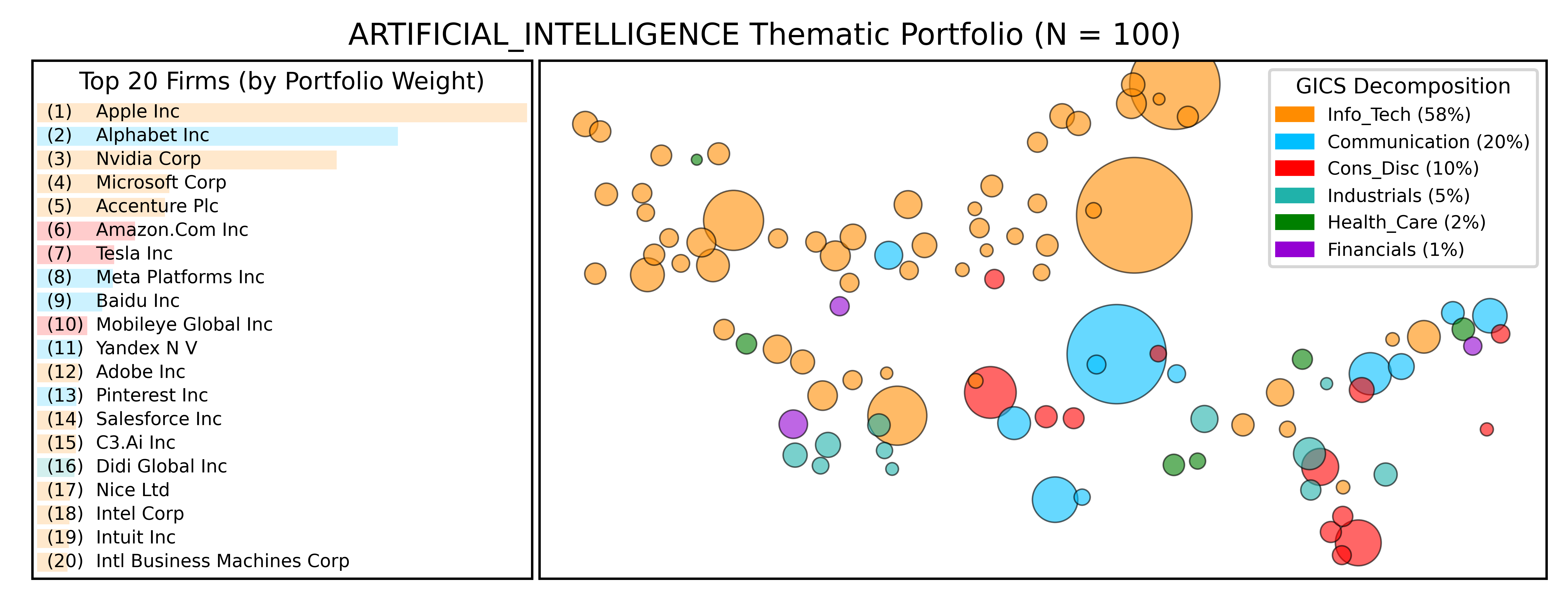}
    \caption{Extraction of AI Firms from Relevant GICS Sectors}
\end{figure}
\vspace{-0.5cm}

The ten largest firms represent three sectors, highlighting just how sector-agnostic \emph{artificial intelligence} really is. Beyond that, the long tail of smaller names also includes firms in industrials, healthcare, and financials - producing a portfolio that would not be possible to construct systematically using GICS. 

As a second example, consider a \emph{robotics} portfolio : 
\vspace{-0.3cm}
\begin{figure}[h!]
    \centering
    \includegraphics[width=16cm]{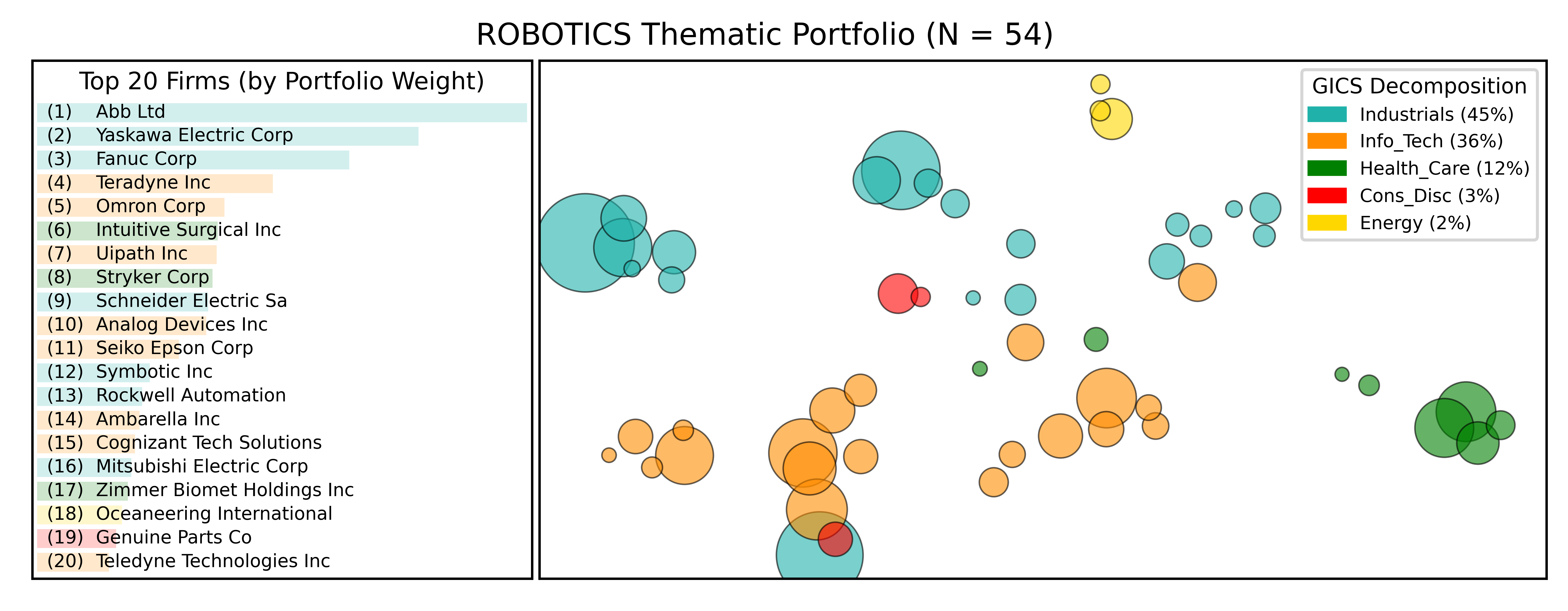}
    \caption{Extraction of AI Firms from Relevant GICS Sectors}
\end{figure}
\vspace{-0.6cm}

As above, this portfolio represents multiple GICS sectors, including a cluster of surgical robots (healthcare), oil drilling robots (energy), industrial robots, and several firms in information technology. Without a tool like MIS, the only way to construct such a portfolio would be to do it entirely by hand, thus highlighting the significant utility of our model to investors who want to allocate to emerging industries. 

\newpage
\section{Conclusion}
What we present here is a proof-of-concept. Probabilistic industry classification is still a highly nascent area of research, with many exciting potential improvements and useful applications. MIS is merely a starting point for a much larger conversation on the various applications of big data and advanced artificial intelligence systems for improving the quality of services that asset managers can provide. 

\subsection{Current Model Limitations}
In the spirit of intellectual honesty and academic transparency, we believe that it is critically important for us to address key limitations of our process as we iterate towards better solutions : 
\vspace{-0.3cm}
\begin{itemize}
    \item \textbf{Bias :} as mentioned in section 2.3, MIS involves text pre-processing via the construction of semantic trees, which must be done by hand and thus reflects the subjective perspectives of the practitioner.
    \item \textbf{Incompleteness :} as described in section 3.2, MIS leverages a mixed-membership model that categorizes the individual words of the underlying text. Thus, the model is only able to identify industries that are explicitly mentioned in the input text data.  
    \item \textbf{Noise :} as demonstrated in section 3.4, MIS utilizes posterior approximation for estimating industry-mixtures, which introduces variation due to randomness in the optimization process. 
\end{itemize}
\vspace{-0.3cm}
Despite such limitations, which are unfortunately unavoidable for many artificial intelligence pipelines, we find that MIS estimated industry-mixtures are incredibly useful. Further, when sufficient care is applied to data sourcing, text pre-processing, and model regularization - the negative effects of these limitations can be significantly mitigated, providing a reliable tool for asset management applications. 

\subsection{Future Research Directions}

As with any emerging technology, there exist many obvious low-hanging-fruit for additional investigation. Here, we highlight three potential extensions that we are particularly excited about : 
\vspace{-0.3cm}
\begin{itemize}
    \item \textbf{LLM Text Pre-Processing :} though we have been critical of LLMs as the primary machinery for industry classification, we believe that there exist many interesting ways that they can be used for low-impact tasks such as text pre-processing, in which they may help to de-noise unruly documents.  
    \item \textbf{Private Equity Proxies :} beyond simply identifying the nearest neighbors of a publicly traded firm, we can use MIS to identify industry profiles for privately held firms, using only a text-based description. This allows for approximate risk modeling for firms that do not provide quantitative data publicly. 
    \item \textbf{Risk Modeling :} many popular fundamental risk models use GICS to stratify systematic risk by industry. By replacing GICS with MIS it may be possible to improve the quality of fit for such models. 
\end{itemize}
\vspace{-0.3cm}
We intend to explore each of these direction of research in the coming years, and hope to inspire industry peers to develop innovations in similar technologies. 

\subsection{Closing Remarks}

Though GICS has proven immensely useful for asset managers over the past several decades, the advent of the artificial intelligence has paved the way for more robust data-oriented approaches that can circumvent the limitations of existing systems. Amazon is so much more than a bookstore, and MIS provides a rigorous framework to describe it. 

\newpage
\bibliographystyle{ieeetr}
\bibliography{main}

\end{document}